
\def\({\left(}
\def\){\right)}
\def\[{\left[}
\def\]{\right]}

\def\a{\alpha}
\def\be{\beta}
\def\g{\gamma}
\def\G{\Gamma}
\def\de{\delta}

\def\la{\lambda}

\def\th{\theta}
\def\om{\omega}
\def\Om{\Omega}

\def\s{\sigma}
\def\S{\Sigma}

\def\coma{\quad ,\quad}
\def\build#1_#2^#3{\mathrel{\mathop{\kern
																			0pt#1}\limits_{#2}^{#3}}}

\def\frac#1#2{{#1\over#2}}

\def\sqr#1#2{{\vcenter{\vbox{\hrule height.#2pt
									\hbox{\vrule width.#2pt height#1pt \kern#1pt
            \vrule width.#2pt}
         \hrule height.#2pt}}}}

\def\vspace#1{\crcr\noalign{\vskip#1\relax}}

\def\vs#1{\vspace{#1mm}}

\magnification 1200
\overfullrule 2pt

{\leftskip 3mm
\rightskip 3mm
\noindent{\bf AN APPROXIMATE  GLOBAL SOLUTION TO THE GRAVITATIONAL FIELD OF A
PERFECT FLUID IN SLOW ROTATION}
\footnote{*}{A translation from  Spanish of the paper  ``Soluci\'on aproximada al
campo gravitatorio de un fluido perfecto en rotaci\'on lenta'', {\it Encuentros
Relativistas Espa\~noles\/}, Publicaciones del IAC, Serie C - No. 6, 1988; pp.
215-230} \par}
\bigskip
\centerline{J. A. Cabezas and E. Ruiz}
\centerline{Grupo de F\'\i{}sica  Te\'orica, Universidad de Salamanca, 37008
Salamanca, Spain}
\vskip 10mm
\centerline{\bf Abstract}

{\leftskip 15mm
\rightskip 15mm\noindent
Using the Post--Minkowskian  formalism and considering rotation as a perturbation,
we compute an approximate interior solution for a stationary perfect fluid with
constant density and axial symmetry. A suitable change of coordinates allows this
metric to be matched to the exterior metric to a particle with a
pole-dipole-quadrupole structure, relating the parameters of both.\par}

\vskip 15mm
\noindent{\bf 1. Introduction.}
\medskip

In the study of multipole  particles in General Relativity, certain approximate
solutions to the gravitational field outside such objects have been proposed. The
information about their structure is collected in a series of parameters, which are
interpreted as the multipole moments of the particle. Here, we use as a starting
point the metric outside a particle with pole-dipole-quadrupole structure (PDQ
particle), with stationary and axisymmetric moments [1]. Our aim is to find an
approximate interior metric that will match it on a given surface. It is demanded
that they should match in the sense of Lichnerowicz [2]: the metric and its first
derivatives must be continuous on the matching surface. For simplicity's sake we
shall assume that the inside is made up of a homogeneous perfect fluid in rigid and
slow rotation (this term will be precised below). The interior solution will be
obtained by solving the Einstein equations using the Post--Minkowskian perturbation
method [3]. From study of the motion equations of the material medium it is
possible to extract precise information for obtaining the equation of the matching
surface [4]. Finally, the matching conditions of both metrics will be translated
into relationships between the parameters of the interior metric (density and
angular speed of rotation), the mean radius of the surface, and the parameters of
the exterior metric (multipole moments).
\bigskip

\noindent{\bf 2. The exterior metric.}
\medskip

In a system of  harmonic  coordinates $X^\alpha$ such that $X^0=t$ and $\phi$ are
adapted to the Killings vectors ($R$, $\Theta$, and $\phi$ are the spherical
coordinates associated to the harmonic ones), the exterior metric to an isolated
PDQ particle with a stationary and axisymmetric structure is written as follows
(see ref. 1):
$$
\eqalign{
g_{00}^{\rm ext} =&-1 
+2\frac{Gm}{Rc^2}\(1-\frac{Gm}{Rc^2}+\frac{G^2m^2}{R^2c^4}-\frac{G^3m^3}{R^3c^6}\) 
+2\frac{GB}{R^3c^2}P_2(\cos\Theta) \cr
\vs2
& -\frac{G^2}{R^4c^4}
\[4mB\,P_2(\cos\Theta)-\frac{2S^2}{3c^2}\[1+2P_2(\cos\Theta)\]\] + O(R^{-5})\cr
\vs2
g_{0i}^{\rm ext} =&  2\frac{GS}{R^2 c^3}\(1-\frac{Gn}{Rc^2}+\frac{G^2m^2}{R^2
c^4}\)M_i + O(R^{-5})\cr
\vs2
g_ {ij}^{\rm ext} =& 
\[1+2\frac{Gm}{Rc^2}+\frac{G^2m^2}{R^2c^4}+2\frac{GB}{R^3c^2}P_2 \right.\cr
\vs2
&\left.\qquad- \frac{G^2}{R^4c^4}\Big[\frac{mB}3(1-7P_2)
+\frac{S^2}{c^2}(2-3P_2)\Big]\]\de_{ij} \cr
\vs2
&+ \[\frac{G^2m^2}{R^2c^4}\Big[1+2\frac{Gm}{Rc^2}+2\frac{G^2m^2}{R^2c^4}\Big]
\right.
\vs2
&\left.\qquad+ \frac{G^2}{R^2c^4}\Big[mB(2+5P_2)+\frac{S^2}{c^2}(4+3P_2)\Big]\] N_i
N_j \cr
\vs2
&+\frac{G^2}{R^4c^4}\[mB+4\frac{S^2}{c^2}\] E_i
E_j-2\frac{G^2}{R^4c^4}\[3mB+5\frac{S^2}{c^2}\]\cos\Theta\,N_{(i}E_{j)} \cr}
\eqno(2.1)$$
where
$$
E^i= \de^i_3 \coma N^i = \frac{X^i}{R} \coma M^i =
\epsilon^{i}_{\ jk} N^j E^k
\eqno(2.2)$$
$P_l(\cos\Theta)$ is  the Legendre polynomial of order $l$ and the parameters $m$,
$S$ and $B$, are  respectively the mass,  the angular momentum and the mass
quadrupole moment of the particle.

Note that the terms  containing the different moments  have a specific dependence
on the powers of $R^{-1}$. More precisely, it may be seen that only the following
factors appear:
$$
\frac{Gm}{Rc^2} \coma \frac{GS}{R^2 c^3} \coma 
\frac{GB}{R^3 c^2}
\eqno(2.3)$$
and their powers and products.

Finally, it is  worth mentioning that (2.1) coincides  with the expansions, in
certain harmonic coordinates, of  Schwarzschild, Weyl and Kerr metrics, if the
parameters $m$, $S$ and $B$ take the values:
$$
\matrix{
Gm=\a \hfill& GS=0 \hfill& GB = 0 \hfill&{\rm(Schwarzschild)}\cr
\vs2
Gm=-a_0 \hfill& GS=0 \hfill& GB = \frac13 a_0^3-a_2 \hfill&{\rm(Weyl)}\cr
\vs2
Gm=\a \hfill& GS=-\a\,a \hfill& GB = -\a\,a^2 \hfill&{\rm(Kerr)}\cr
}
\eqno(2.4)$$
\bigskip

\noindent{\bf 3. Newtonian theory.}
\medskip

The Post--Minkowskian  approximation allows  one to obtain the interior metric in
the form of an expansion in powers of G, which we shall match, at each iteration,
with the terms of the same order of the exterior solution. Since the results
obtained at first--order must necessarily coincide with those afforded by Newtonian
mechanics, it makes sense to first study the problem posed within this framework.

It should be recalled  that study  of a self-gravitating fluid is already a
non--linear problem in Newtonian mechanics. On one hand, the field equation:
$$
\triangle\psi_G = 4\pi G\rho
\eqno(3.1)$$
allows one to obtain  the gravitational  potential, $\psi_G$, once we know the mass
distribution $\rho$. On the other hand, Euler motion equations:
$$
\eqalign{
&\partial_t\rho + \partial_k(\rho v^k) = 0 \cr
\vs2
&\partial_t(\rho v^i) + \partial_k(\rho v^k v^i +\s^{ki}) =-\rho\partial^i\psi_G \cr
}
\eqno(3.2)$$
(where $v^i$ is  velocity and $\sigma^{ik}$ is  the stress tensor of the fluid)
afford information about the evolution of the source of the gravitational field, if
we have prior knowledge of the potential $\psi_G$ inside it.

Let us assume that the  fluid  is perfect ($\sigma^{ik}=p\delta^{ik}$), that
density is constant in time ($\partial_t\rho=0$) and that the fluid rotates in a
rigid manner with a constant angular velocity $\omega^i$ ($v^i=\epsilon^i_{\
jk}\omega^jX^k$). Under these conditions, equation (3.2) can be reduced to (we
choose the $X^3$ axis in the direction of $\omega^i$):
$$
\partial_\phi\rho =0 \coma \partial_i p = -\rho\partial_i\psi
\eqno(3.3)$$
with
$$
\psi = \psi_G -\frac12 \om^2 r^2 \sin^2\Theta
\eqno(3.4)$$
The second equation in (3.3) is  a  constraint on $p$, $\rho$ and $\psi$. If one
considers this as an equation for $p$, from its integrability condition
$$
\partial_{[i}\,\rho\partial_{j]}\psi = 0
\eqno(3.5)$$
it may be deduced that $\rho$ must be  a  function of $\psi$, $\rho=\chi(\psi)$ (or
the other way round); and its solution is:
$$
p= p_0 - \int_{\psi_0}^\psi \chi(\la)\,d\la
\eqno(3.6)$$
The surface $\Sigma$ of the fluid is   determined by the equilibrium condition
$p\mid_\Sigma=0$, which leads to
$$
p= \int_\psi^{\psi_\S} \chi(\la)\,d\la
\eqno(3.7)$$
where $\psi_\Sigma$ is  the constant  value of $\psi$ on $\Sigma$ (hence,
$\psi(R,\Theta)=\psi_\Sigma$ is the implicit equation of
$\Sigma$). Moreover, since  $p$ and $\rho$ are functions of
$\psi$, they are constant on the surfaces $\psi(R,\Theta)={\rm constant}$.

If an equation of state $p=f(\rho)$  is  given as an input, (3.3) allows one to
obtain a relation  $\rho=\chi(\psi)$, which, substituted in (3.1), leads to a
generally non--linear equation for the potential:
$$
\triangle\psi_G = 4\pi\,\chi(\psi)
\eqno(3.8)$$
For example, for a polytropic--like state equation 
$$
p= a\,\rho^{1+\frac1{\g}} \qquad (a={\rm constant}\ ,\ \g>0)
\eqno(3.9)$$
one obtains
$$
\rho = C\,(\psi_\S-\psi)^\g \coma C=
\(\frac{\g-1}{a\,\g}\)^{\!\g}
\eqno(3.10)$$
and
$$
\triangle\psi_G = 4\pi G\,C\,(\psi_\S-\psi)^\g
\eqno(3.11)$$
which is not linear if $\g\neq 1$ [5].

The case of $\rho={\rm constant}$, which is not included in the previous example,
 is not even free of non--linearities. We have already mentioned that
$\psi(R,\Theta)=\psi_\Sigma$ defines the surface $\Sigma$. However, to determine it
we must first know the potential $\psi_G$, which in turns depends on the shape of
the surface
$$
\psi_G = -G\int_{({\rm interior\ of} \ \Sigma)}\ \frac{\rho}{\mid\vec X-\vec
X'\mid}d^3\vec X'
\eqno(3.12)$$
Accordingly, $\psi=\psi_\Sigma$, rather than a definition of the surface,  is a
supplementary condition that must satisfy the solution.

Although different exact  solutions to this  problem are known [6], it can also be
posed in terms of successive approximations. In the strict sense, this second
procedure reproduces one of the known exact solutions (the McLaurin ellipsoids).
Nevertheless, it is highly suitable for the problem we shall study in later
sections.

As is well known, the expansion of (3.12)  in spherical harmonics in the region
exterior to $\Sigma$ is:
$$
\psi_G^{\rm ext} = -\frac{Gm}{R} - \sum_{l\ge
2}\frac{G\,Q_l}{R^{l+1}}P_l(\cos\Theta)
\eqno(3.13)$$
where
$$
Q_l \equiv \int_{({\rm interior\ of}\ \Sigma)}\ R^l P_l(\cos\th)\,\rho(\vec R\,)
\,d^3\!\vec R
\eqno(3.14)$$
is the $2^l$-pole moment. We shall admit that there is a parameter $\Omega$
($\Omega\ll 1$), related to the rotation, that measures the deformation of the
fluid with respect to the spherical shape, such that $Q_l$ is at least of order
$\Omega^l$. Furthermore, we shall  assume that it is possible to expand the
parametric function of the surface of the fluid, $R_\Sigma(\cos\Theta)$
($\psi\bigl(R_\Sigma(\cos\Theta),\cos\Theta\bigr)=\psi_\Sigma$), in powers of
$\Omega$. Direct, although tedious, calculation based on expressions (3.13) and
(3.14) and the equation for $\Sigma$ leads to the following results:
$$
\eqalign{
&R_\S(\cos\th) = R_0\[1-\frac56\Om^2P_2 \right.\cr
\vs2
&\phantom{espacio largo largo}\left.+ \Om^4\(-\frac5{36}P_0-\frac{50}{63}P_2 +
\frac{15}{14}P_4\)+ O(\Om^6)\] \cr
\vs2
&Q_2 = mR_0^2\[-\frac12\Om^2-\frac5{21}\Om^4+O(\Om^6)\]\cr
\vs2
&Q_4 = mR_0^4\[\frac{15}{28}\Om^4+O(\Om^6)\]\cr
}
\eqno(3.15)$$
At the end of the process, it  is possible to identify the parameter $\Omega$,
which proves to be:
$$
\Om^2 = \frac{R_0^2 \om^2}{Gm/R_0} = 
\frac{R_0 \om^2}{Gm/R_0^2}
\eqno(3.16)$$
which is the quotient between  the rotational  energy and the gravitational energy,
or, between the mean values of the centrifugal force and the gravitational force on
the surface.
\bigskip

\noindent{\bf 4. Relativistic Euler equations.}
\medskip

Within the framework  of relativity, it is  also possible to gain information about
the matching surface from the evolution equations of the fluid. These equations are
deduced from the energy--momentum tensor conservation condition, which, for a
perfect fluid:
$$
T^{\a\be} = \(\rho + \frac{p}{c^2}\)u^\a u^\be +
\frac{p}{c^2}g^{\a\be} \qquad (u^\a u_\a =-1)
\eqno(4.1)$$
can be written
$$
\eqalign{
&u^\a\partial_\a\rho + \(\rho + \frac{p}{c^2}\)\partial_\a u^\a = 0 \cr
\vs2
&\(\rho + \frac{p}{c^2}\)u^\be\partial_\be u^\a +
\frac1{c^2}(g^{\a\be}+u^\a u^\be)\partial_\be \rho =0 \cr
}
\eqno(4.2)$$
If the metric is stationary  and axysimmetric  and, also, the velocity of the fluid
is in the plane spanned by both Killing vectors, it is possible to find coordinates
adapted to the Killing vectors in which
$$
u^\a = \psi\,\(\de^\a_t + \frac{\om}{c}\de^\a_\phi\)
\eqno(4.3)$$
and the metric is written in  blocks [7]. From the condition $u^\alpha
u_\alpha=-1$, one deduces that
$$
\psi = \[-\(g_{tt}+2\frac{\om}{c}g_{t\phi}+\frac{\om^2}{c^2}
g_{\phi\phi}\)\]^{-\frac12}
\eqno(4.4)$$

Under the hypothesis that  rotation is rigid ($\omega={\rm constant}$) and that
density and pressure do not depend on either $t$ or $\phi$, the second set of Euler
equations is reduced to
$$
\partial_a p = (\rho c^2 + p))\partial_a\ln\psi
\qquad (a,b,\dots = R\,,\Theta)
\eqno(4.5)$$
Considering this generalisation of the second equation of (3.3) as an
equation for the pressure, the integrability condition leads us to
$\rho=\chi(\psi)$, and the solution is:
$$
p = \int_{\psi_\S}^\psi\frac{\chi(\la)}{\la^2}d\la
\eqno(4.6)$$
As in the Newtonian case, $\rho$ and $p$ are  constant over the surfaces
$\psi(R,\cos\Theta)={\rm constant}$, and $\Sigma$ (the surface on which pressure
vanishes) has as its implicit equation $\psi(R,\cos\Theta)=\psi_\Sigma$. If the
fluid considered has a constant density, (4.6) leads to:
$$
p = \rho\(\frac{\psi}{\psi_\S}-1\)
\eqno(4.7)$$
\bigskip

\noindent{\bf 5. Parameters of the exterior metric.}
\medskip

As mentioned in the introduction, the exterior  metric  that we shall use as the
starting point is a linear combination of products of factors (2.3) up to $R^{-4}$.
Additionally, the interior metric will be calculated by the Post--Minkowskian
approximation, considering at each order in $G$ a subexpansion in powers of
$\Omega$. To match them to order $G$, it is of interest to know how the terms (2.3)
are written as a function of these parameters. Instead of $G$, we shall use the
adimensional parameter $g=Gm/R_0c^2$ ($R_0$ is the mean radius of the ball of
fluid) and we shall consider that the moments have an expansion in $\Omega$ similar
to that obtained in Newtonian mechanics; that is:
$$
\eqalign{
&Q_2 \equiv B = mR_0^2\Big[\Om^2B_{(2)} + \Om^4B_{(4)}+O(\Om^6)\Big] \cr
\vs2
&S= mR_0\om\Big[S_{(0)} + \Om^2S_{(2)}+O(\Om^4)\Big] \cr
}
\eqno(5.1)$$
One thus has:
$$
\eqalign{
&\frac{Gm}{Rc^2} = g\frac{R_0}{R} \cr
\vs2
&\frac{GS}{R^2c^3} = g\frac{S}{mR_0c}\(\frac{R_0}{R}\)^2
=g^{\frac32}\Om\Big[S_{(0)} + O(\Om^2)\Big]\(\frac{R_0}{R}\)^2\cr
\vs2
&\frac{GB}{R^3c^2} = g\frac{B}{mR_0^2}\(\frac{R_0}{R}\)^3  =g\Big[\Om^2B_{(2)} +
O(\Om^4)\Big]\(\frac{R_0}{R}\)^3\cr }
\eqno(5.2)$$
We thereby manage  to substitute the expansion in  powers of $R^{-1}$ (which lacks
sense for the interior metric) by an expansion in $g$ and $\Omega$. This new
approach leads to a different  assignation of orders to the terms of the expansion
of the exterior metric. For example, according to the first criterion
$G^2mB/R^4c^6$ and $G^2S^2/R^4c^6$ are $G^2/R^4$ terms. However, now, whereas the
first  is of order $g^2\Omega^2$, the second proves to be of order $g^3\Omega^2$.
This is a direct consequence, in  sum, of having chosen $\Omega$ instead of another
parameter as an estimate of the effect of rotation. Nevertheless, this is the only
way to recover the Newtonian results at order $g$.

Let us now see the meaning of making expansions in these parameters. The parameter
$g$ is the quotient between the Schwarzschild radius of the body ($R_s=Gm/c^2$) and
$R_0$. Therefore, $g$ will be small when $R_0\gg R_s$. On the other hand, $\Omega$
has a strange relationship with $g$:
$$
\Om^2 = \frac1{g}\(\frac{R_0\om}{c}\)^2
\eqno(5.3)$$
If the body is not very compact (small $g$), the typical velocity with which a
point at the surface moves ($R_0\Omega$) must be very small as compared with the
speed of light, so that a small value of $\Omega$ can be attained and so that the
approximation will conserve its validity. By contrast, when the body is compact,
the surface velocity can be an important  fraction of the speed of light. In the
case of the Earth, the Sun and a typical pulsar ($m_{\rm pulsar}\simeq m_\odot$,
$T\simeq 1\,{\rm s}$, $R_0\simeq 10\,{\rm km}$), $g$ and
$\Omega$ take the following values:
$$
\matrix{
\ &{\rm Earth} & {\rm Sun} & {\rm pulsar} \cr
\vs2
\hfill g\,: & 1.3\times 10^{-9} & 2.1\times 10^{-6}& 0.14 \cr
\vs2
\hfill\Om^2\,: &1.8\times 10^{-3} &3.3\times 10^{-5} & 2.9\times
10^{-7} \cr
}
\eqno(5.4)$$

Finally, we can use the expansion in these parameters to determine the matching
surface in an approximate way from the condition $\psi(R,\cos\Theta)=\psi_\Sigma$ as
a function of  the parameters of the exterior metric (we assume continuity of the
metric on $\Sigma$). The final result is:
$$
\eqalign{
R_\S(\cos\th) = R_0\[1+ \Om^2\(B_{(2)}-\frac13 +
\frac43g\Big[S_{(0)}-1\Big] +O(g^2)\)P_2\]+ O(\Om^4) \cr
}
\eqno(5.5)$$
It should be noted that since the exterior metric does not contain the octupole
moment metric (if we assume equatorial symmetry) nor the 16-pole moment, which would
go as $\Omega^4$, there is little sense in considering expansions beyond $\Omega^2$.
\bigskip

\noindent{\bf 6. The interior metric}
\medskip

The Post--Minkowskian perturbation method presupposes the existence of a formal
expansion in the gravitational constant $G$ of $h^{\alpha\beta}\equiv  \bigr(-{\rm
det}(g_{\gamma\mu})\bigl)^{1/2}g^{\alpha\beta}-\eta^{\alpha\beta}$ [8]. In our case,
since in the exterior metric there are half--integer powers of $g$, we  shall assume
that the expansion is in half--integer powers of $g$ (an adimensional parameter).
$$
h^{\a\be}(x^\mu) = \sum_{n\ge 2}g^{\frac{n}2}
\,h^{\a\be}_{(n/2)}(x^\mu,\Om)
\eqno(6.1)$$
Although later, as mentioned above, we subexpand  in powers of $\Omega$ each of the
terms of this expansion, the structure of (6.1)  ensures that we can use the whole
Post--Newtonian scheme, whose details, together with the explicit expressions for
the development of the different magnitudes appearing in the Einstein equations, can
be found in the papers cited in reference 7. Accordingly, they will not be referred
to here.

With a view to simplifying the problem  as much  as possible, we shall assume that
the perfect fluid is homogeneous ($\rho={\rm constant}$), in which case pressure
will be given by (4.7). Since $\psi\simeq 1+O(g)$ pressure is order $g$, which means
that the energy-momentum tensor at zero--order will be expressed only in terms of
$\rho$ (Newtonian theory).
\medskip

{\it Order $g$.} The solution of the Einstein  equations at this order, which is
regular at the origin, is
$$
\eqalign{
&h_{(1)}^{00} =  2\frac{\rho'}{m}\(\frac{r}{R_0}\)^2 +\sum_{n\ge
0}H_n^{(1)}\(\frac{r}{R_0}\)^n P_n \cr
\vs2
&h_{(1)}^{0i}=h_{(1)}^{ij}={\rm solution\ of\ the\ homogeneous\ equation} \cr
}
\eqno(6.2)$$
with $\rho^\prime\equiv (4\pi/3)\rho R_0^3$. The fact that the parameter of the
exterior metric $m$ appears in (6.2) arises from having chosen $g$ as the constant
in which we expand the solution, and has no further transcendence.

Our aim is to match this metric with (2.1) in the sense of Lichnerowicz (see ref.
2): the metric and its first derivatives should be continuous on the surface
determined by (5.5). For the exterior metric (2.1), $h_{{\rm ext}\,(1)}^{0i}=h_{{\rm
ext}\,(1)}^{ij}=0$; we thus choose the solution of the homogeneous equation
mentioned in (6.2) equal to zero. Also, the fact that in the exterior metric only
terms in $P_0$ and $P_2$ appear implies that, by continuity, $H_n^{(1)}=0$ if $n>2$.
Since  there are four constants, $m$, $B$, $H_0^{(1)}$ and $H_2^{(1)}$, and since
four conditions are imposed by the  matching (terms in $P_0$ and
$P_2$ in $h_{(1)}^{00}$ and their normal derivative to $\Sigma$), the matching of the
two metrics is possible  by identifying the exterior $X^\alpha$ and interior
$x^\beta$ coordinates. Thus,
$X^\a=x^\a + O(g^{3/2})$ and the metrics match if:
$$
\eqalign{
&m = \rho' + O(g) \cr
\vs2
& B = -\frac12 \rho'R_0^2\Om^2\Big[1+ O(g)\Big] \cr
\vs2
& R_\S = R_0\[ 1-\frac56\Om^2 P_2 + O(g\Om^2) + O(\Om^4)\] \cr
}
\eqno(6.3)$$
Finally, one has that
$$
h_{(1)}^{00} = 2\(\frac{r}{R_0}\)^{\!\!2} -6 +
2\Om^2\(\frac{r}{R_0}\)^{\!\!2} P_2 + O(\Om^4)
\eqno(6.4)$$
\medskip

{\it Order $g^{3/2}$.} Calculation of the solution at this order passes through the
same steps as the previous order and in a completely similar way, since only the
$h_{(3/2)}^{0i}$ components are different from zero and, in fact, depend on a single
function. Neither is it necessary here to modify the interior or exterior
coordinates; that is $X^\a=x^\a + O(g^2)$. The matching conditions demand that
$$
S= \frac25\rho'R_0^2\om\Big[1+ O(\Om^2) +O(g)\Big]
\eqno(6.5)$$
Finally,
$$
\eqalign{
&h_{(3/2)}^{0i} = \[2\frac{r}{R_0} -\frac65\(\frac{r}{R_0}\)^3 +O(\Om^2)\]\Om m^i \cr
\vs2
&h_{(3/2)}^{00} =h_{(3/2)}^{ij} = 0 \cr
}
\eqno(6.6)$$
\medskip

{\it Order $g^2$.} At this order, the solution  obtained
$$
\eqalign{
h_{(2)}^{00} =\ &  \(\frac{\rho'}{m}\)^2\[18\(\frac{r}{R_0}\)^2 -
\frac52\(\frac{r}{R_0}\)^4\] +\Om^2\frac{\rho'}{m}\(\frac25-\frac{37}7
P_2\)\(\frac{r}{R_0}\)^4 \cr
\vs2
&+ \sum_{n\ge 0}H_n^{(2)}\(\frac{r}{R_0}\)^n P_n + O(\Om^4) \cr
h_{(2)}^{0i} =\ &{\rm solution\ of\ the\ homogeneous\ equation}\cr
h_{(2)}^{ij} =\ & \[\(\frac{\rho'}{m}\)^{\!\!2}\[\frac65\(\frac{r}{R_0}\)^{\!\!2} 
-\frac37\(\frac{r}{R_0}\)^{\!\!4}\] \right.\cr
\vs2
&\left.+\Om^2\frac{\rho'}{m}
\[\frac25\(\frac{r}{R_0}\)^{\!\!2}+
\frac1{126}(89-203P_2)\(\frac{r}{R_0}\)^{\!\!4}\]\]\de^{ij} \cr
\vs2
&+\[\(\frac{\rho'}{m}\)^{\!\!2} \[-\frac35\(\frac{r}{R_0}\)^{\!\!2} 
+\frac27\(\frac{r}{R_0}\)^{\!\!4}\]
\right.\cr
\vs2
&\left.+\Om^2\frac{\rho'}{m}
\[\frac25\(\frac{r}{R_0}\)^{\!\!2}+  
\(-\frac{67}{63}+\frac5{18}P_2\)
\(\frac{r}{R_0}\)^{\!\!4}\]\] n^i n^j \cr
\vs2
&-\frac56\Om^2\frac{\rho'}{m}\(\frac{r}{R_0}\)^{\!\!4} e^i e^j  
+\frac73\Om^2\frac{\rho'}{m}
\(\frac{r}{R_0}\)^{\!\!4}\!\cos\th\, e^{(i}n^{j)}\cr
\vs2
&+{\rm divergence\ free\ solution\ of\ the\ homogeneous\ equation} + O(\Om^4) \cr
 }
\eqno(6.7)$$
(where $e^i=\delta_3^i$, $n^i=x^i/r$ y $m^i=\epsilon^i_{\ jk}n^je^k$) does not match  the part
corresponding to the exterior metric if one does not make a change of coordinates;
this even occurs if $\Omega=0$ (Schwarzschild).

The coordinate change
$$
X^\a = x^\a -g^2\,f^\a(X^\mu) + O(g^{5/2})
\eqno(6.8)$$
induces the change
$$
h'^{\a\be} = h^{\a\be} +g^2\Big[\eta^{\a\be}\partial_\mu
f^\mu -2\partial^{(\a}f^{\be)}\Big] + O(g^{5/2})
\eqno(6.9)$$
in the metric density. In order  to preserve the structure and explicit symmetry of
the metric, one must take:
$$
\eqalign{&f^0 = 0 \cr
\vs2
&f^i = \sum_{l\ge 1}f_l^i = \sum_{l\ge 1}  
\[\chi_l(R)\frac{P_l^1}{\sin\th}(N^i-\cos\th\,E^i) +\G_l(R)P_l\,E^i\]\cr }
\eqno(6.10)$$
($P_l^1$ are the associated Legendre functions  of order 1) where $\chi_l$ and
$\Gamma_l(R)$ are arbitrary functions. The new coordinates $X^\a$ are still harmonic,
for the interior metric, if $\triangle f=0$. It  is possible to check that the
change in the spatial components of the metric density that induces the most general
solution of this equation, which is regular at the origin of coordinates, has the
same structure as the solution of the same type  and divergence free of the
homogeneous equation that must be satisfied by the spatial components of
$h_{(2)}^{ij}$. Accordingly, we can take  this  term out of (6.7) since it is
included in the change of coordinates.

In order to not introduce into the metric polynomials in $\cos\Theta$ of a higher
degree than necessary, only $f_1^i$ and $f_3^j$ must be taken into account, with the
restriction
$$
3\chi_3 + \G_3 = 0
\eqno(6.11)$$

The $h_{(2)}^{0i}$ component, which  is not modified by  the coordinate change, must
be zero in order to match the exterior. Also, the matching of $h_{(2)}^{00}$ implies
four conditions and that of $h_{(2)}^{ij}$ twelve (the term $\delta^{ij}$ provides
four, $n^in^j$, four, $n^{(i}e^{j)}$, two and $e^ie^j$ two more). As free parameters,
one has $H_0^{(2)}$, $H_2^{(2)}$, $m$ and $B$ [by the latter two we understand the
corrections of order $g$ to the Newtonian values (6.3)], which allow us to match
$h_{(2)}^{00}$ with and without a  coordinate change (see order $g$), and also the
parameters provided by the coordinate change.

The change (6.10) with the restrictions indicated introduces  three independent
arbitrary functions $\chi_1$, $\Gamma_1$ and $\chi_3$. If $\Omega=0$ (spherical
symmetry), (6.9) reduces to $f^i=f(R)N^i$; that is, a single arbitrary function that
is a particular combination of $\chi_1$ and $\Gamma_1$. Therefore, all  the
functions, with the exception of this combination, are at least of order $\Omega^2$.
The constants introduced by the coordinate change are the values of these functions
and their first and second derivatives on the surface $\Sigma$. At the order of
approximation required, this is equivalent, for  functions  of order $\Omega^2$, to
taking the value of the function (and respectively its derivatives) in $R_0$; for
example:
$$
\chi_3\mid_\S = \chi_3(R_0) +O(\Om^4)
\eqno(6.12)$$
However, for the combination that we have called $f$,
we must conserve the first two terms of its expansion in powers of $\Omega$
$$
f\mid_\S = f(R_0) - \frac56\Om^2f'(R_0)\,P_2 + O(\Om^4)
\eqno(6.13)$$
and the same for its derivatives. To conclude, we have available $4+3+3=10$ arbitrary
constants (actually, nine  because under matching conditions two of them always
appear in the same combination) to satisfy twelve conditions. As a result, three of
them must be linear combinations of the rest. This is the case here.

Having achieved the continuity of the $h_{(2)}^{ij}$ components and their
derivatives, and therefore having determined the constants associated with the
coordinate change, we look at the $h_{(2)}^{00}$ component. From the matching
we obtain:
$$
\eqalign{
&m = \rho'\[1+g\Big[3+\frac25\Om^2 + O(\Om^4)\Big] +O(g^2)\] \cr
\vs2
&B = \rho'R_0^2\[\Om^2\Big[-\frac12+\(\frac{1682}{105}g+ O(g^2)\) \Big] + O(\Om^4) \] \cr
}
\eqno(6.14)$$

Finally, it should be mentioned that by changing the  coordinates not only of the interior but also of
the exterior metric one can find a global harmonic coordinate system; i.e., in these
coordinates the metric and its first derivatives are continuous on $\Sigma$.
\vskip 20mm

[1] J. Mart\'\i{}n and E. Ruiz, Phys.  Rev.  {\bf D32}, 2550 (1985); Y. G\"ursel,
Gen. Rel. Grav. {\bf 15}, 737 (1983).

[2] A. Lichnerowicz,  {\it Th\'eories Relativistes  de la Gravitation et de
l'\'Electromagn\'e\-tisme\/} (Mason, Paris, 1955).

[3] Referring to the method  we use as Post--Minkowskian, this is  perharps
misleading. In fact, a true Post--Minkowskian approximation should exhibit
explicit Minkows\-kian covariance. This is not the case in our approach, since we use
definite systems of coordinates all along the calculations (``mass--centered''
coordinate systems).

[4] R. H. Boyer, Proc. Camb. Phil. Soc. {\bf 61}, 527 (1965).

[5] It is possible to obtain  for $\rho$ an equation like (3.11). See for instance,
S. Chandrasekar, {\it Stellar Structure\/} (Dover, New York, 1939).

[6] J. P. Luminet, Ann. Phys. Fr.  {\bf 10}, 101 (1985); C. Y. Wong, Astrophys. J.
{\bf 190}, 675 (1974).

[7] D. Kramer, H. Stephani, M. MacCallum   and H. Herlt, {\it Exact Solutions of
Einstein's Field Equation\/}  (Cambridge University Press, Cambridge, 1980).

[8] L. Bel, T. Damour, N. Deruelle,  J. Iba\~nez  and J. Mart\'\i{}n, Gen. Rel.
Grav. {\bf 13}, 963 (1981); the second paper cited in ref. 1.

\bye